# Fitting the Galactic Electron Spectrum Measured by Voyager at Low Energies and also by CALET/AMS-2 at High Energies Using a Monte Carlo Diffusion Model for Electron Propagation - One Spectrum Fits All

**W.R. Webber[1], J. Quenby[2] T. Harrison[1], S. Stochaj[3], and T.L. Villa[1]**

1. New Mexico State University, Astronomy Department, Las Cruces, NM 88003, USA

2. Blackett Laboratory, Imperial College, London, SW7 2BZ UK

3. New Mexico State University, Physics Department, Las Cruces, NM 88003, USA

# ABSTRACT

In this paper we compare galactic electron spectra measured up to TeV energies by AMS-2 and CALET and that measured by Voyager at the lowest energies down to 1 MeV with that calculated using a Monte Carlo diffusion model for electron propagation in the galaxy. The observations and calculations at both ends of the electron spectrum, differing by a factor ~$10^6$ in energy, can be matched to within a few percent with a minimal set of assumptions. This includes an electron spectrum of the form $\left(\frac{dj}{dE}\right) \sim E^{-2.1\,(1+m\log E)}$ which has an index increasing from ~2.1 at the lowest energies to 2.4 at ~1 TeV, along with a diffusion coefficient that remains essentially constant below ~1.5 GV above which it becomes ~$P^{0.45}$. The Monte Carlo calculations indicate that the lowest and highest energy electrons originate within a local region near the galactic plane of size, $\leq$ L where L is the thickness of the trapping region. The remarkable agreement to within a few percent between the calculations and measurements from 20 GeV up to ~1 TeV also indicates that there are no individual sources >10-20% of the Monte Carlo calculated background that are contributing to the observed electron intensity in this energy range.

**1. Introduction**

The Galactic cosmic ray electron spectra measured by CALET (Adriani, et al., 2018) and by AMS-2 (Aguilar, et al., 2014) set a new standard in the precision measurement of these electrons at high energies. The AMS-2 electron spectrum is fully compatible at the ~3% level with that from CALET but does not yet extend above 1 TeV. This high level of precision is extended to a limit of almost 5 TeV by the CALET measurements. The measurements at low energies by Voyager (Cummings, et al., 2016) and these high energy spectra may be used to describe the distribution in the galaxy of these electrons that we observe at the Earth.

In an earlier paper, as part of a comparison of Voyager data on low energy electrons with higher energy spectrometer measurements, we matched the Voyager measurements with the AMS-2 intensities at 10 GeV using a Monte Carlo diffusion model for the electron propagation between a few MeV to ~10 GeV (Webber and Villa, 2017).

The object of this paper is to extend these calculations to ~10 TeV and use them to interpret existing galactic cosmic ray electron data from 1 MeV to several TeV.

**2. The Electron Propagation Model – Comparison with CALET Observations**

This Monte Carlo model, first described by Webber and Rockstroh, 1997, and used extensively since, assumes a uniform electron source at a galactic latitude of Z=0. This model has the characteristics of a simple Leaky Box Model (LBM) which works so well in describing the nuclei intensities and measurements by Voyager (Webber, 2018). The model includes all loss processes for electrons as a function of energy and a function of Z, perpendicular to the disk. The average B field at Z=0 is taken to be 6μG. For the calculations here the nominal electron source spectrum is taken to be ~$P^{-2.25}$. The reasons for this average value and the dependence of the diffusion coefficient on rigidity used in the paper will be discussed below.

In a diffusion picture this model may be described as a layered diffusion model in which the dominant field configuration is along layers of flux tubes, one above the other above the galactic disk. Within each layer the mean field is parallel to the galactic disk and follows a spiral configuration.

The loss terms for synchrotron and inverse Compton radiation, which are both ~$E^{2.0}$, are dominant in the propagation calculation above a few GeV as is evident from the Monte Carlo derived spectrum presented along with the Voyager, AMS-2 and CALET measurements in Figures 1 and 2. Above ~0.5 GeV the spectral index of the derived electron spectrum increases rapidly even though the source spectrum does not change. The derived spectrum below ~0.5 GeV is mainly influenced by the electron source spectrum and the rigidity dependence of the diffusion coefficient.

Above ~10 GeV the electron source spectrum coupled with the synchrotron and inverse Compton losses and diffusion losses from a diffusion coefficient which is ~$E^{0.45}$, lead to a calculated spectrum with an index that has a slope ~3.15 above ~30 GeV which is consistent with both AMS-2 and CALET measurements. This slope continues over a factor ~50 in energy up to ~1.5 TeV. Between 1.5 and 7.5 TeV the calculated electron spectrum then steepens noticeably to an average exponent ~4.2.

At low rigidities, below ~0.3 GV, the measured electron spectrum flattens noticeably to a spectral index ~1.35 in the Voyager energy range from 3-60 MeV. In order to match this spectral index there are two extreme possibilities: One is that the diffusion coefficient remains constant at low rigidities and as a result the source spectral index must decreases from -2.25 to -2.1 below ~0.15 GV. Alternately the $P^{-2.25}$ spectral index at higher rigidities remains constant down to the lowest rigidities and as a result the diffusion coefficient becomes ~$P^{0.8}$, thus producing the $E^{-1.35}$ spectral index observed at Voyager. Both of these possibilities fit the data and are shown in Figures 1 and 3.

Figure 3 shows the overall dependence of the diffusion coefficient used in the Monte Carlo calculations. It is expressed in Steps To The Boundary (STB), $L/\ell$, where L = boundary distance and $\ell$ = step size, and thus $\ell = \frac{2K}{\upsilon}$, where K is the diffusion coefficient and $\upsilon$ = c for electrons.

Above ~0.3 GV it is seen from Figure 3 that the diffusion coefficient remains constant up to about 1.76 GV before becoming ~$P^{0.45}$ up to the highest rigidities. This high rigidity dependence is based on the AMS-2 measurements of the B/C ratio (Aguilar, et al., 2016) and the interpretation of those results using a Leaky Box propagation model with appropriate cross sections into B (Webber and Villa, 2017).

The reason for the flat region with the largest STB or smallest diffusion coefficient which extends between 0.3 and 1.7 GV, is the observation by Voyager of a high intensity of cosmic ray secondary nuclei produced during interstellar propagation. In particular the $^3$He secondary production is consistent with that from a constant production path length ~9-10 g/cm$^3$ of interstellar H and He from ~3-100 MeV/nuc in the galaxy (Webber, et al., 2018). This large matter path length requires a small and nearly constant diffusion coefficient over the corresponding rigidity range.

## 3. Further Discussion

### 3a. The Low Energy Electron Spectrum

The low energy part of the electron spectrum is defined by the Voyager measurements of a $E^{-1.35}$ spectrum from 3 to 60 MeV. If one accepts that the source spectrum at the lower rigidities is a continuation of the $P^{-2.25}$ spectrum at higher rigidities then the Voyager data can be explained as follows: (1) The diffusion coefficient would need to become ~$P^{0.8-0.9}$ down to the lowest rigidities. Such a possibility has, in fact, been suggested by Ptuskin, et al., 2006. In this case, as described by Webber and Villa, 2017, the low energy electrons would stream out of the galaxy creating a dark halo containing mostly dark energy. The LIS electrons we actually measure would come from distances in the galactic plane that are less than L, the distance to the boundary.

(2) If instead we consider the other alternative, which will also produce the spectrum seen at Voyager, and which is simply a constant diffusion coefficient extended to the lowest rigidities; then the source spectrum itself would need to flatten to be ~$P^{-2.1}$ below ~0.3 GV. This alternative, with the STB remaining constant below 0.3 GV is shown as blue lines in Figures 1 and 3.

### 3b. The High Energy Electron Spectrum

The CALET and AMS-2 data and the Monte Carlo calculations both give an electron spectral index =-3.15 from ~30 GeV to ~1.5 TeV. The absolute values of the measured and calculated intensities agree to within a few percent throughout the entire energy region as seen in Figure 2. The calculations show that this is achieved with a source spectrum with an index increasing from -2.25 to -2.40, which is steepened to an index ~-3.15 by synchrotron and inverse Compton energy losses and also by diffusion escape losses due to the fact that the diffusion coefficient increases ~$P^{0.45}$ at high rigidities as determined from B/C observations up to ~1 TV.

Except for one energy region as noted later, significant differences greater than ± 3% in the Monte Carlo calculated intensities and the AMS-2 or CALET measured electron intensities are not apparent between ~20 GeV and 1-2 TeV. Above ~1-2 TeV the Monte Carlo calculated electron spectral index begins to increase from an index = -3.15 and reaches an average index above ~1.5 TeV. There are indications that an increase in spectral index is also seen by CALET.

The diffusion life time in the Monte Carlo Model is equal to $\tau = L^2/2K$. In our model with L=1 Kpc and K=2 x $10^{28}$ cm$^2$/s at 1 GV the lifetime becomes = 2 x $10^7$ yr at 1 GV, which is in agreement with the value of 1.5 x $10^7$ yr obtained using $^{10}$Be decay measurements (Yanasuk, et al, 2001). At 1 TeV the diffusion lifetime becomes only ~$10^6$ year. Also at 1 TeV the synchrotron and inverse Compton loss lifetimes are even smaller. The lifetimes are both so short that, along the galactic disk, these losses influence the "observed" spectral index at distances from the source which are ~L or greater as seen from the increasing spectral index obtained from the calculations shown in Figure 2.

## 4. Summary and Conclusions

A comparison of galactic electron intensities that are measured by AMS-2 and CALET at high energies and Voyager at low energies and also the intensities calculated using a Monte Carlo diffusion program provides a sensitive insight on the overall propagation and spectrum of these electrons in the galaxy. The observations and calculations can be matched at both ends of the energy spectrum from a few MeV to a few TeV to within a few percent. This "matching" is achieved with minimal assumptions regarding the rigidity dependence of the diffusion coefficient and the electron source spectrum over this broad energy range and is a powerful tool.

In one example considered in this paper the diffusion coefficient is assumed to be independent of rigidity below ~1.76 GV (blue line in Figure 3). Above that rigidity, K ~$P^{0.45}$ as based on AMS-2 B/C measurements. In this case the spectrum for electrons is ~$P^{-2.1}$ at about 0.1

GV rigidity and increases slowly to P$^{-2.40}$ at the highest rigidities. A formula for the electron spectrum where $\left(\frac{dJ}{dE}\right) \sim E^{-2.1\,(1+m\log E)}$ where m = 0.05 will fit the data over ~6 magnitudes of energy. Energy dependent spectral indices with a slowly increasing spectral index are present in the cosmic ray spectra obtained in the acceleration models of Ellison, Jones and Reynolds 1990, for example.

This Monte Carlo description for propagation and source spectra, including what are essentially the same propagation and source parameters that have also been used in Leaky Box calculations which explain the observed spectra of galactic protons and with only slight changes to the log part of the source spectral index, to describe the spectra of He and heavier nuclei as well (see Webber, 2018).

The Monte Carlo calculations indicate that, at both the lowest and highest energies, the electrons we measure originate from a local source region of size <L where L is the ⊥ scale of the diffusing region. For electrons coming from larger distances, at the highest energies, the effects of the b·E$^2$ energy loss terms from synchrotron and inverse Compton losses cause a steepening electron spectrum above ~1-2 TeV. This affect may be observable in the CALET data.

The fact that the Monte Carlo "diffuse" electron intensities agree with those of both AMS-2 and CALET to within just a few percent over the entire energy range from 30 GeV to almost 1 TeV indicates that there are no significant point sources contributing to the observed electron intensity in this energy range. At ~1 TeV, however, there is an excess ~10-15% in the CALET intensities above the Monte Carlo distribution. The excess has been discussed as a possible "source" by the CALET collaboration.

At low energies it is possible to assume that the electron source spectrum is ~P$^{-2.25}$. This would be a continuation of the spectrum assumed between 0.3 and 1.76 GV. In this case, to produce the P$^{-1.35}$ spectrum observed at Voyager the diffusion coefficient would need to be ~P$^{0.8-0.9}$. As a result of the increasingly larger diffusion coefficient at lower energies these electrons would move rapidly out of the galactic disk, thus creating an unobservable halo population of electrons containing "dark" energy (Webber and Villa, 2017) - a different source of "dark" energy?

**Acknowledgements:** The authors are grateful to the Voyager team that designed and built the CRS experiment with the hope that one day it would measure the galactic spectra of nuclei and electrons. This includes the present team with Ed Stone as PI, Alan Cummings, Nand Lal and Bryant Heikkila, and to others who are no longer members of the team, F.B. McDonald and R.E. Vogt. Their prescience will not be forgotten. This work has been supported throughout the more than 40 years since the launch of the Voyagers by the JPL.

# REFERENCES

Adriani, O., Akaike, Y., Asano, K., et al., 2018, http://arXiv.org/abs/1806.09728

Aguilar, M., Aisa, D., Alpat, B., et al., 2014, Phys. Rev. Lett., 113, 221102

Aguilar, M., Cavasonza, L.A., Ambrosi, G., et al., 2016, Phys. Rev. Lett, 117, 231102

Cummings, A.C., Stone, E.C., Heikkila, B.C., et al., 2016, Ap.J., 831, 21

Ellison, D.C., Jones, F.C. and Reynolds, S.P., 1990, Ap.J., 360, 702

Moskalenko, I.V. and Strong, A.W., 1998, Ap.J., 493, 694-707

Ptuskin, V.S., Moskalenko, I.V., Jones, F.C., et al., 2006, Ap.J., 642, 902

Webber, W.R. and Rockstroh, J.M., 1997, Adv. Space Res., 19, 817-820

Webber, W.R. and Villa, T.L., 2017, http://arXiv.org/abs/1711.08015

Webber, W.R. and Villa, T.L., 2017, http://arXiv.org/abs/1703.10688

Webber, W.R., 2018, Cosmic Rays, Intechopen, 72533, p.61

Yanasuk, N.E., Wiedenbeck. M.E. Mewaldt, R.A., et al., 2001, Ap.J., 563, 768-792

# FIGURE CAPTIONS

**Figure 1:** Electron intensities measured by Voyager (Cummings, et al., 2016) and by AMS-2 (Aguilar, et al., 2014), with the intensities x $E^2$. The AMS-2 intensities are corrected for positrons. The Monte Carlo calculations #164 and 166 (blue line) are shown. The secondary e- contribution to the Monte Carlo spectrum is shown as a shaded region.

**Figure 2:** Electron intensities at higher energies measured by AMS-2 (Aguilar, et al., 2014) and by CALET (Adriani, et al., 2018). The Monte Carlo calculations are shown in black with an additional secondary e- component added as a shaded region below ~30 GV. The plot shows intensity x $E^3$ on a linear scale so intensity differences of only $\pm$ 5% are readily apparent.

**Figure 3:** The diffusion coefficient used in the Monte Carlo calculations shown in terms of STB (Steps to the Boundary) where $\ell$ is the step size and $\ell = \frac{2K}{v}$, where K is the diffusion coefficient.

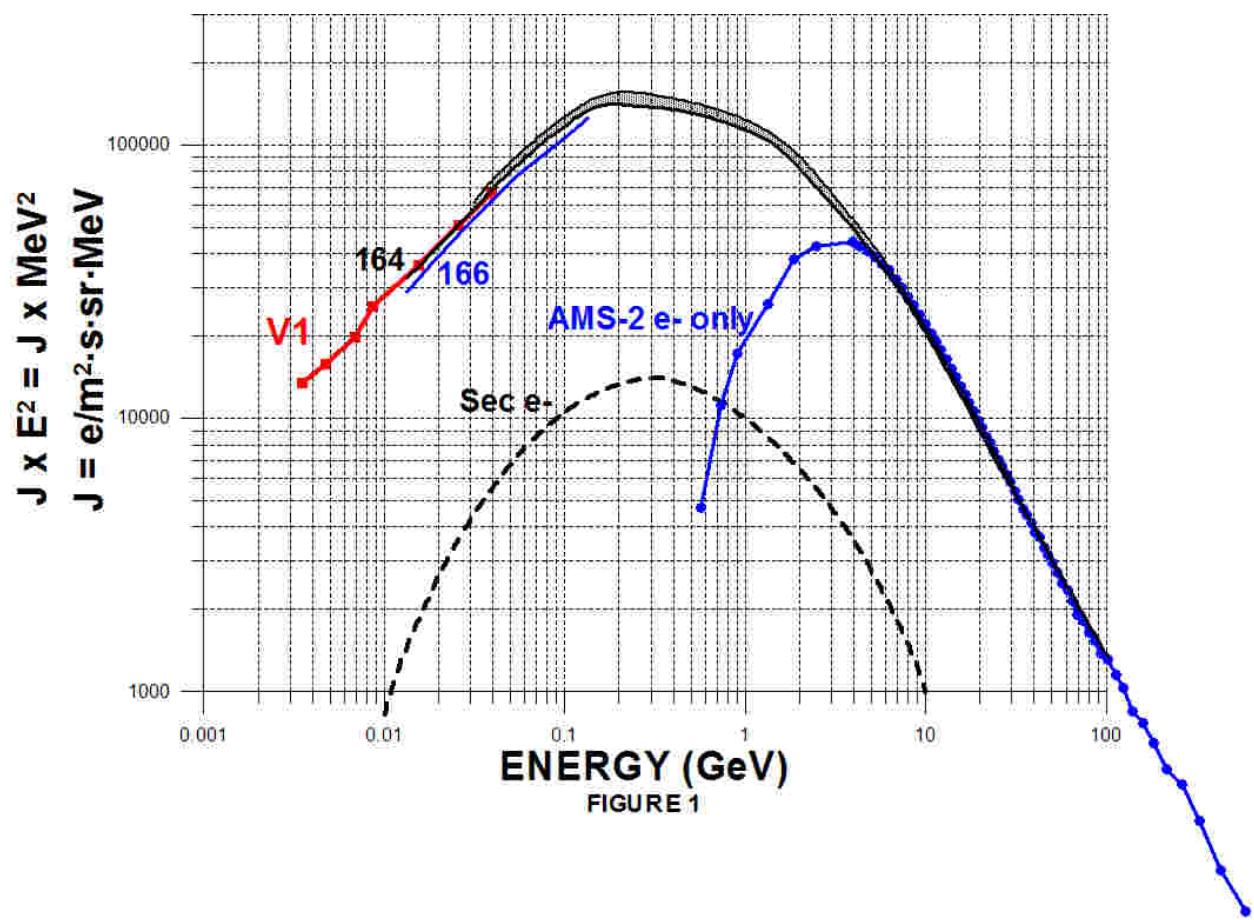

FIGURE 1

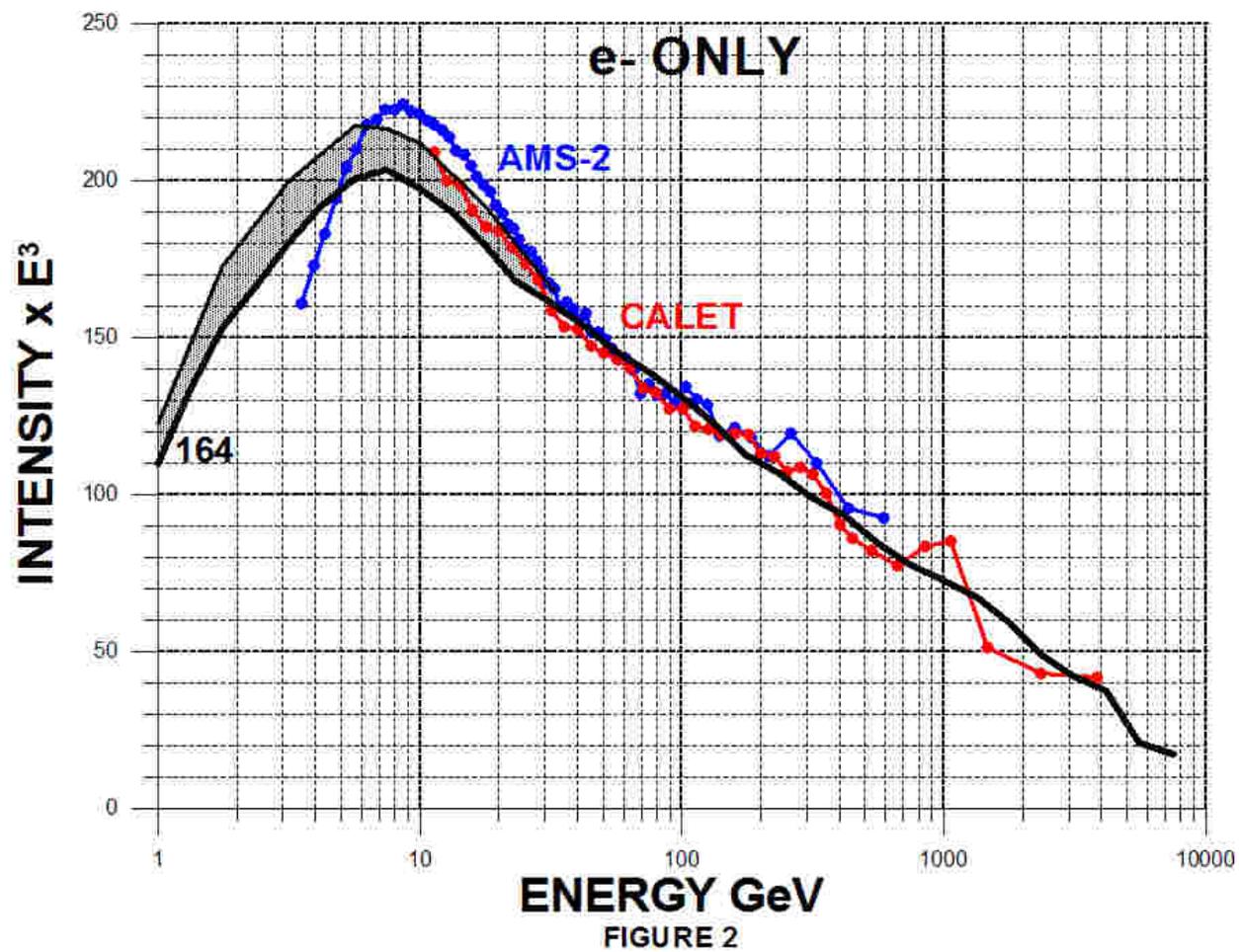

FIGURE 2

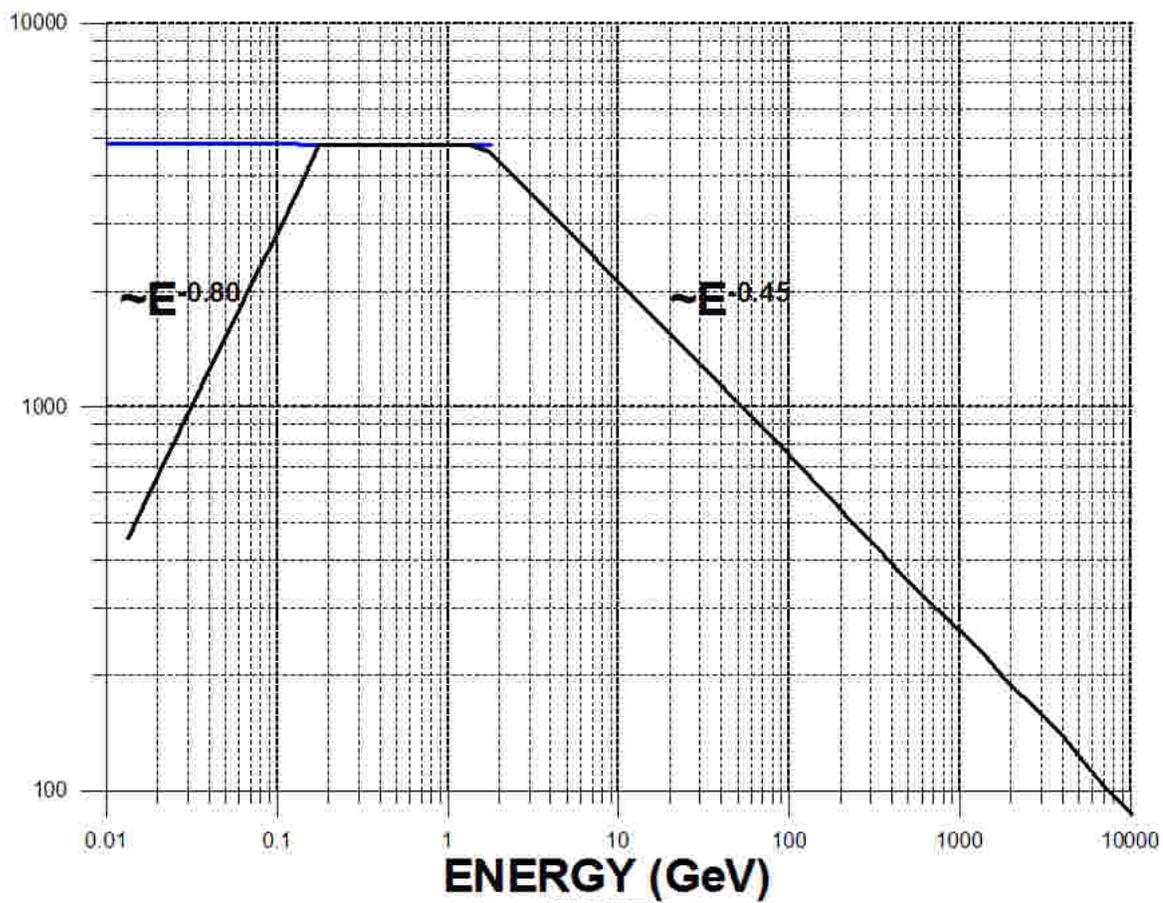

**FIGURE 3**